\documentclass[12pt]{article}
\usepackage{fullpage,amsmath,epsfig}

\newcommand\intvx{\int\!d^5\!x\,}
\newcommand\intivx{\int\!d^4\!x\,}

\begin{document}

\title{On the sign of the dilaton in the soft wall models}

\author{
Andreas Karch$^1$, Emanuel Katz$^2$, Dam T. Son$^3$,
Mikhail A. Stephanov$^4$ \\
{\small $1$: Department of Physics, University of Washington, Seattle, WA
98195} \\
{\small $2$: Department of Physics, Boston University, Boston, MA, 02215}\\
{\small $3$: Institute for Nuclear Theory, University of Washington,
Seattle, Wa 98195}\\
{\small $4$: Department of Physics, University of Illinois, Chicago,
IL 60607}\\
{\small Email:
{\tt karch@phys.washington.edu, amikatz@bu.edu, }} \\
{\tt \small son@phys.washington.edu, misha@uic.edu}
}

\maketitle

\abstract{ We elaborate on the existence of a spurious massless scalar
  mode in the vector channel of soft-wall models with incorrectly
  chosen sign of the exponential profile defining the wall.  We
  re-iterate the point made in our earlier paper and demonstrate that
  the presence of the mode is robust, depending only on the infra-red
  asymptotics of the wall. We also re-emphasize that desired
  confinement properties can be realized with the correct sign
  choice.}

\section{Introduction}
\label{sec:intro}

A few years back we proposed in \cite{Karch:2006pv} the holographic
``soft-wall" model for QCD.  This purely phenomenological model employs
a gauge field $V_{M}$ living in a five dimensional spacetime and
describes an infinite tower of vector meson states in QCD.  The model is
governed by the action
\begin{equation}
  \label{eq:S}
    S = -\frac1{4}\intvx\sqrt {g}\, e^{-\Phi} V^{MN}V_{MN}
\end{equation}
describing a gauge field propagating in the ``background'' in
which not only the metric $g_{\mu\nu}$ but also the dilaton field
$\Phi$ has nontrivial dependence on the 5th coordinate $z$.  Here we
consider a U(1) gauge theory for simplicity (for, e.g., isospin
singlet mesons), so the field strength is given by $V_{MN}
= \partial_M V_N -\partial_N V_M$.  As in all holographic models, this
quadratic action determines the masses of mesons via its mode
spectrum, and gives current/current correlation functions via its
response to a source inserted at the boundary at $z=0$.  It was
observed in \cite{Karch:2006pv} that a pure AdS background metric
\begin{equation}
  \label{gwant}
  ds^2 \equiv g_{MN}\,dx^Mdx^N=z^{-2} \left(dt^2 - d\vec{x}^2 - dz^2\right)
\end{equation}
together with a quadratic scalar profile
\begin{equation}
  \label{pwant}
  \Phi = a z^2
\end{equation}
where $a$ is a {\em positive} constant, leads to the meson spectrum with
squared masses that grow linearly with radial excitation number
$n$, $m_n^2\sim n$ for $n\gg1$, as expected from semiclassical
quantization of a flux tube.  This seems to fit
phenomenology better than the $m_n^2\sim n^2$ growth encountered in
``hard wall'' holographic models.  Generalizing the discussion to
also include fields of
higher spin $S$, the full Regge behavior
\begin{equation}
  \label{regge}
  m_{n,S}^2 = 4 a (n+S) \qquad \mbox{[$a>0$]},
\end{equation}
is reproduced.

Two questions arise when choosing the background.  The
first question is related to the fact that we characterized the
background both by a metric as well as a background scalar
$e^{-\Phi}$, but (in the frame where $g^{zz}=g^{xx}=-g^{tt}$)
only the particular combination
\begin{equation}\label{e-B}
e^{-B} = \sqrt{g}\,(g^{tt})^2\, e^{-\Phi}= {z^{-1}}\,{e^{-az^2}}
\end{equation}
appears in the action for the vector mesons. Does it matter how the
desired factor $e^{-az^2}$ in $e^{-B}$ is split between the metric and
the scalar?  This question was answered in \cite{Karch:2006pv}
by noting that when considering higher spin particles
only a pure
AdS metric~(\ref{gwant})
together with a quadratic scalar profile (\ref{pwant})
gives rise to the desired Regge spectrum (\ref{regge}) not just for
the vector mesons but also for the higher spin fields.

The second question is due to the observation that for vector mesons ($S=1$)
the spectrum~(\ref{regge})
is independent of the sign of $a$. What then dictates the sign
choice for $a$ in eq.(\ref{e-B})? In a footnote in Ref.~\cite{Karch:2006pv},
we argue that the negative sign $a<0$ should be discarded,
because it would give rise
to massless states in the {\em vector} channel.  Such massless
states indicate spontaneous breakdown of the vector symmetry, which does not
happen in QCD.  This argument leaves the
positive sign $a>0$ as the only physically acceptable choice.
There is an additional argument against the $a<0$ choice, not
mentioned in \cite{Karch:2006pv}.  Namely,
the independence of the mass spectrum on the sign of $a$ does
not extend to higher spin mesons. Choosing $a<0$ would give a higher
spin meson spectrum independent of spin,
\begin{equation}
  \label{eq:sgn-a-S}
   m_{n,S}^2 = 4 |a| (n + 1)  \qquad \mbox{[$a<0$]},
\end{equation}
making it unphysical in this respect as well.

Recently the negative sign choice for $a$ in Eq.~(\ref{e-B}), for the
description of $S=1$ mesons, has been revived in several papers
\cite{deTeramond:2009xk,Zuo:2009dz,Nicotri:2010at} as it seemed
to have nicer confinement properties as probed by external test
quarks. The issue of the extra massless particle was either claimed to be
non-existent~\cite{Nicotri:2010at}, regarded
as a peculiarity of
our exactly solvable model that could be easily remedied
\cite{deTeramond:2009xk}, or something that can be eliminated by
a choice of boundary condition~\cite{Zuo:2009dz}.

Given these recent developments, we wish to take the time to elaborate
on our original footnote in Ref.~\cite{Karch:2006pv} concerning the
sign of $a$ in Eq.~(\ref{e-B}) and demonstrate in detail that the
unphysical massless vector meson is present in the negative sign,
$a<0$, dilaton soft wall model. Our main point in
Sec.~\ref{sec:massl-state-vect} is that this massless state is
generic, robust, and can not easily be remedied by boundary
conditions. It remains as a normalizable mode as long as
$\int_0^{\infty} dz\, e^B$ is {\em finite} and so can not be removed
by simply changing details of $e^B$ without changing its
asymptotics. It should also be noted that this issue is independent of
the question of whether $e^B$ is due to a non-trivial metric or scalar
field profile. The existence of the extra massless mode only depends
on the form of $e^B$.

In Sec.~\ref{sec:conf-extern-test} we discuss the potential between
heavy test quarks in the negative sign dilaton soft wall, which was
the main motivation of this alternate model. It was
argued in \cite{Andreev:2006ct} that one would get a
phenomenologically appealing heavy quark potential with a short range
Coulomb potential and a long range linear potential from a minimal
area calculation in a background of the form eq.(\ref{e-B}) with the
negative sign $a<0$ as long as the non-trivial $e^B$ is entirely due
to the {\it metric}. We already mentioned above that the case with the
exponential factor carried by the metric does not reproduce Regge
trajectories for mesons of higher spin when taken as the basis of a
full soft-wall model.  In Sec.~\ref{sec:conf-extern-test} we further
argue that the choice of the negative sign $a<0$ background is both
unnatural and unnecessary as far as confining properties of the model
are concerned.

\section{The massless state in the vector meson channel}
\label{sec:massl-state-vect}

As our main reason to discard the positive sign soft-wall model was
the appearance of an extra massless state in the vector meson
channel. This point has been called into question in different ways in
\cite{deTeramond:2009xk,Zuo:2009dz,Nicotri:2010at}. Here we discuss in
detail the appearance of this mode in the negative sign $a<0$ soft
wall models. We give three alternative derivations, any one of which
demonstrates the existence of this unphysical mode. The three different
approaches shed light on different aspects of the problem and better
address different counterarguments.

\subsection{The vector mode equation}
The equations of motion for the vector mesons following from the
action Eq.~(\ref{eq:S}),
\begin{equation}
  \label{eq:eomV}
  \partial_M\left(\sqrt g \, V^{MN}\right)=0,
\end{equation}
are most easily analyzed in ``axial'' $V_5=0$ gauge, even though, as we
shall see in Section~\ref{sec:wilson}, the physical origin of the massless mode is easier to understand in a different gauge.
For the 4d-transverse components $V_{\mu}(z,\vec{x},t)=V_{\mu}^T(\vec{x},t)\, v(z)$ (with $\partial^{\mu}
V_{\mu}(\vec{x},t)^T=0$) the normalizable modes $v_n$ have to obey
\begin{equation} \partial_z (e^{-B} \partial_z v_n ) + m_n^2 e^{-B} v_n =0 \end{equation}
which after changing variables as
\begin{equation}
\label{veom}
v_n = e^{B/2} \psi_n  = {e^{a z^2/2}}{\sqrt{z}}\, \psi_n \end{equation}
translates into the Sch\"odinger like equation
\begin{equation}
\label{rho}
- \psi_n'' + \left (a^2 z^2 + \frac{3}{4z^2} \right ) \psi_n = m_n^2 \psi_n.
\end{equation}
As pointed out in Ref.\cite{Karch:2006pv} and re-emphasized recently in \cite{Nicotri:2010at}, this equation is
independent of the sign choice for $a$ in (\ref{e-B}), so naively the vector meson spectrum is independent of whether one
discusses the positive or negative sign soft wall. One should keep in mind however that the spectrum is only determined by the
equations of motion together with a set of boundary conditions, which here come from the condition that the fluctuation be
normalizable. The proper quantity to consider is the on-shell action evaluated on the solution to (\ref{rho}). The on-shell
action has two terms with different $z$ dependence. The 4d ``kinetic" term proportional to $V_{\mu \nu} V^{\mu \nu}$ has a
prefactor
proportional to
\begin{equation}
N_k = \int dz \, e^{-B} (v_n)^2 = \int dz \, (\psi_n)^2.
\end{equation}
Requiring this to remain finite reduces to the standard Schr\"odinger norm. The 4d mass term for these massive vector fields
proportional to $m_n^2 V_{\mu} V^{\mu}$ has a prefactor
\begin{equation}
N_m = \int dz \, e^{-B} (v_n')^2.
\end{equation}
For $m_n \neq 0$ this expression can be integrated by parts to yield, using eq.(\ref{veom}), $N_m = m_n^2 N_k$.
So using the standard Schr\"odinger norm on $\psi$ gives the correct
spectrum at non-zero mass. For zero mass one however has to be a
little more careful. For genuine vector modes with $V_{\mu \nu} \neq
0$ one needs to require $N_k$ and hence the Schr\"odinger norm to be
finite. As eq.(\ref{rho}) does not allow any normalizable {\em zero}
energy solutions, there are indeed no massless genuine vector modes in
the system.

However, the 5d vector potential can also propagate 4d {\it scalar} modes, which in the axial gauge $V_5=0$ we have been using
take the form
\begin{equation}
V^S_{\mu} = F(z) \, \partial_{\mu} \pi(\vec{x},t) .
\end{equation}
For a mode of this form $V_{\mu \nu}$ vanishes identically as partial derivatives commute. So finiteness of $N_k$ is not an
issue as now $N_m$ plays the role of the kinetic term. For $m_n \neq 0$ we found before that $N_m$ and $N_k$ are proportional to
each other, so no new normalizable modes of this form exist there either. However for $m_n=0$ we find that
\begin{equation}
\label{zeromode} V^S_{\mu} =F(z) \, \partial_{\mu} \pi(\vec{x},t)
\quad\mbox{with}\quad F'(z) = C e^{B(z)}
\quad\mbox{and}\quad \partial_\mu\partial^\mu\pi(\vec{x},t)=0,
\end{equation}
is a solution to the equation of motion (\ref{eq:eomV}). This solution is
normalizable (has finite $N_m$) as long as
\begin{equation}
N_m = \int_0^\infty dz \, e^{-B} (F')^2
= C^2 (\partial_{\mu} \pi)^2 \, \int_0^\infty dz \, e^B
\end{equation}
is a finite integral. We hence see that the relevant condition that
ensures existence of this extra massless scalar mode is finiteness of
$I\equiv \int dz e^B$. The integrand $e^B$ vanishes as $z$ in the UV
(at $z=0$), so the integral is UV finite. In the positive sign (that is $a>0$)
soft
wall model $e^B$ diverges exponentially in the IR. So $I$ is divergent and
there is no massless scalar mode. In the negative sign ($a<0$) soft wall
however $e^B$ goes to zero exponentially, $I$ is a finite integral and
the extra massless mode is present. In order to confirm this fact, we
shall explicitly demonstrate the corresponding $1/q^2$ pole in the
current/current 2-point function in the next subsection and also give
a more physical interpretation of the extra massless mode as arising
from fluctuations of the phase of the Wilson line (in a certain gauge
which will be specified).
For now let us note that the finiteness of $I$ is a generic feature of any
model with the same asymptotic behavior of $B$ and not just
a peculiarity of a particular choice of $B$ as suggested in
\cite{deTeramond:2009xk}.

\subsection{The pole in the current/current correlation function}

As we have pointed out in the last subsection, the meson spectrum
contains a massless scalar mode. This extra scalar mode should make
its presence known as a $1/q^2$ pole in the vector-current-to-vector-current
correlation function
\begin{equation}
  \label{eq:JJ}
  \intivx\,e^{iqx}\,\langle J^\mu(x) J^\nu(0)\rangle
= \left(q^\mu q^\nu - q^2\eta^{\mu\nu}\right)\Pi_V(-q^2).
\end{equation}
In this respect it is clearly distinct from the
physical pion field, which lives in the {\em axial} current sector. To
determine the 2-pt function we need to solve eq.~(\ref{veom}) with
$m_n^2$ replaced by  $q^2$.
Unlike the normalizable modes $v_n$ of the discrete spectrum, a
non-normalizable solution $v(q,z)$ exists for any $q^2$, and can be
chosen to satisfy $v(q,z)=1$.  In addition, we want the solution
$v(q,z)$ to obey the same boundary condition in the IR as the modes
$v_n$, i.e., remain bounded. For the $a<0$ soft wall the solution has
been nicely discussed in \cite{Zuo:2009dz}, and for the $a>0$ in,
e.g., \cite{Zuo:2008re}. Here we shall discuss both cases in
parallel. The solution to the Schr\"odinger equation~(\ref{rho}) with
$m_n^2$ replaced by $q^2$, bounded as $z\to\infty$ is given by
\begin{equation}
  \label{eq:psi-q2}
  \psi(q,z) = \frac{\rm const}{\sqrt z}  e^{-|a|z^2/2}
     U\left(\frac{-q^2}{4|a|},0,|a|z^2\right)
\end{equation}
where $U$ is the Tricomi confluent hypergeometric function. Another
solution, obtained by changing the sign in front of $|a|$ in
Eq.~(\ref{eq:psi-q2}) is exponentially divergent as $z\to\infty$.
The solution $v(q,z)$ that we need is then given by
\begin{equation}
\label{currentsolution}
v(q,z) = \sqrt{z} e^{az^2/2} \psi(q,z) =
\Gamma\left(1-\frac{q^2}{4|a|}\right)\, e^{(a-|a|)z^2/2}\,
   U\left(\frac{-q^2}{4|a|},0,|a|z^2\right)
\end{equation}
From eq.(\ref{currentsolution}) we can extract the 2-pt function as usual
\begin{equation}\label{eq:PiV}
\Pi_V(-q^2) = \left . \frac{e^{-B}\partial_z v(q,z)}{q^2} \right |_{z=\epsilon}
=  \left .  \frac{\partial_z v(q,z)}{q^2z} \right |_{z=\epsilon}
= \frac{a-|a|}{q^2} - \frac{1}{2} \psi\left(1-\frac{q^2}{4|a|}\right)
  + {\rm const}.
\end{equation}
where $\psi$ denotes the digamma function, with poles located at positive
values of $q^2=m_n^2=4|a|(n+1)$. The $1/q^2$ pole for $a<0$
clearly signals the presence of a massless, propagating
mode. Ref. \cite{Nicotri:2010at} tries to argue that this is a contact
term and hence can be canceled by a suitable counterterm. Contact
terms are UV singularities and indeed can be regulated, but they are
polynomial in $q^2$. The $1/q^2$ pole is an IR singularity and
hence contains real physics: the appearance of an extra massless
state.

As the massive vectors satisfy the same mode equation for either sign
of $a$, we have the same eigenvalues and eigenfunctions and so their
contribution sums up into the same digamma function. This particular
form of the current correlation function had been earlier conjectured
in e.g. \cite{Shifman:2000jv,Cata:2006ak} based on a model of the
spectrum processed via dispersion relations. The only difference
compared to the $a<0$ soft wall is that the extra pole due to the
massless scalar in the vector channel is absent. This explicitly
confirms our earlier analysis of the modes.

\subsection{The Wilson line representation}
\label{sec:wilson}

We exhibited the extra massless mode alluded to in our footnote in
\cite{Karch:2006pv} both by an explicit mode analysis as well as a
calculation of the 2-pt function. The physical origin of this mode can
be understood a little more clearly by redoing the mode analysis in a
slightly different gauge. From the higher dimensional point of view the extra
scalar can be naturally thought of as the Wilson line of the gauge
field in the extra dimension. So not surprisingly, a gauge where
$V_5$ is set to zero from the very beginning hides some of the
interesting aspects of the physics. In order to motivate our new gauge
choice, let us expand out the original action in Eq.~(\ref{eq:S}):
\begin{equation}
  \label{eq:S2}
    S 
= \intvx e^{-B}\left[
      -\frac12  (\partial_5 V_\mu - \partial_\mu V_5)^2
- \frac14 V^{\mu\nu}V_{\mu\nu}\right]
\end{equation}
We do want to keep both $V_{\mu}$ and $V_5$ to capture vector and scalar degrees of freedom. This association is complicated by
the presence of the term mixing the two
\begin{equation}
  \label{eq:mix}
  S_{\rm mix}=\intvx e^{-B} (\partial_\mu V_5) (\partial_5 V^\mu)
\end{equation}
In order to simplify this mixing term we, following \cite{Son:2003et}, gauge transform $V_5$ to
\begin{equation}
  \label{eq:A-z-gauge}
  V_5 = C\pi(x) e^{B}
\end{equation}
($C$ is a constant, which we shall choose later by imposing a
convenient normalization condition on the terms quadratic in $\pi$).
The mixing term becomes a total derivative, integrating to
\begin{equation}
  \label{eq:int-mix}
  S_{\rm mix} =  C\intivx
\partial_\mu \pi(x) \left[V^\mu\right(x,z)]_{z=0}^{\infty}
\end{equation}
which is zero if $V_\mu$ vanishes at $z=0$ and $\infty$.

Next, let us calculate the variation $\delta S$ of the (extremal) action under
an infinitesimal variation of the boundary condition $V_\mu(x,z=0)$
around $V_\mu(x,0)=0$ and $\pi\neq0$. This variation
$\delta S$ is coming entirely from the boundary term in
Eq.~(\ref{eq:int-mix}) which gives:
\begin{equation}
  \label{eq:delta-S}
  \delta S =   C\intivx
\,\partial_\mu \pi(x)\, \delta V^\mu(x,0)\,.
\end{equation}
This, by holography and the definition of the Noether current, means:
\begin{equation}
  \label{eq:current}
  J^\mu = C \partial^\mu \pi\,.
\end{equation}
If the field $\pi$ is normalized canonically, i.e., the action is
$S=(1/2)\int d^4x (\partial_\mu\pi)^2+\ldots$, then $C=f_\pi$, by definition
of $f_\pi$. In other words, the current-current correlator will have a
pole
\begin{equation}
  \label{eq:J-J}
  \langle J^\mu J^\nu\rangle_q = -f_\pi^2 q^\mu q^\nu\frac1{q^2}+ \ldots
\end{equation}

Finally, the action for $\pi$ is, from Eqs.~(\ref{eq:S2}) and~(\ref{eq:A-z-gauge})
\begin{equation}
  \label{eq:pi-action}
  S = \frac12 \intvx  e^{-B}\,
\left[C \partial_\mu \pi(x) e^{B}\right]^2
= \frac{C^2}{2}
\left(\int_0^\infty dz\,e^B  \right)
\intivx (\partial_\mu \pi)^2
\end{equation}
which means
\begin{equation}
  \label{eq:f-pi}
  f_\pi^{-2}=C^{-2}=\int_0^\infty\! dz\,e^B
\end{equation}
As long as $f_\pi$ is finite, the extra massless mode in $V_5$ will
represent a genuinely new massless degree of freedom that contributes
to the correlation functions, as we've already seen
above. Furthermore,
for $e^B$ given by Eq.~(\ref{e-B}), evaluating integral in
Eq.~(\ref{eq:f-pi}) with $a<0$ gives $f_\pi^2=2|a|$, which is in agreement with
Eqs.~(\ref{eq:PiV}) and~(\ref{eq:J-J}).

Of course, we can always perform a gauge transformation parametrized by a gauge parameter $\lambda(z,\vec{x},t) = C
\pi(\vec{x},t) \int_{0}^z e^{B(\tilde{z})} d\tilde{z}$ to set $V_5=0$ starting from eq.(\ref{eq:A-z-gauge}). This will turn on
\begin{equation} V_{\mu} = - \partial_{\mu} \lambda =  - C \left ( \int_{0}^z d \tilde{z} \, e^{B(\tilde{z})} \right ) \,
\partial_{\mu} \pi(\vec{x},t) \end{equation}
which is indeed exactly the massless mode we identified previously in eq.(\ref{zeromode}) starting in axial gauge directly.

\section{Confinement for external test quarks}
\label{sec:conf-extern-test}

\newcommand\tens{{\cal T}} 

One of the main motivations for re-introducing the negative sign soft
wall despite its obvious shortcoming of having an extra unphysical
massless mode were its seemingly superior confinement properties
\cite{Andreev:2006ct}. Following the standard AdS/CFT procedure
\cite{Maldacena:1998im,Rey:1998ik}, the introduction of the external
static test quarks is described in the bulk by introducing a
fundamental string with its endpoints following the worldlines of the
test quarks out on the boundary. The free energy of the
quark/anti-quark pair can be read off from the on shell action of the
fundamental string. Confinement would occur if, for
large quark-antiquark pair separation, the minimal energy string has a
long segment parallel to the boundary at constant value of
$z=z_*$. This happens if the position-dependent string tension for a
$z={\rm const}$ string, $\tens(z)$, has a minimum at some $z=z_*$, and
$\tens(z_*)> 0$~\cite{Sonnenschein:2000qm}. If the background for the
string propagation is purely geometrical, then the action of the
string is proportional to its proper area and $\tens(z)={\rm
  const}\cdot g_{tt}(z)$ (in the frame with $g_{zz}=g_{xx}=-g_{tt}$). It was observed in \cite{Andreev:2006ct} that for
a purely geometric soft-wall model, with $\Phi=0$ but
$g_{tt}=z^{-2}e^{-2az^2}$ in Eq.~(\ref{e-B}),
the ``potential'' $\tens(z)$ has a minimum only for {\em negative} values of
$a$.

We wish to point out several flaws in the above argument for
justification of the choice $a<0$. First of all, as we emphasized is
Section~\ref{sec:intro}, the soft wall model, in order to describe the
Regge spectrum for higher spin states, must not have any $e^{{\rm const}\cdot z^2}$
factor in the metric. One might imagine also that if the metric was
given as pure AdS
in the Einstein frame, it might acquire $e^{{\rm const}\cdot z^2}$ from the
rescaling to the string frame. However, this would also be in
contradiction with our  analysis of the higher spin states in
Ref.~\cite{Karch:2006pv}, which rested on the assumption that the
metric is pure AdS in the {\em string} frame.

Second, the confinement criterion~\cite{Sonnenschein:2000qm} we
described above does not require the
$z$-dependent string tension $\tens(z)$ to diverge at infinity. A simple
minimum suffices. The absents of such a minimum in the simple
background we used is clearly not a generic feature. It is reasonable
to assume that a model, where the dilaton background is generated
dynamically, will have a nontrivial metric profile as well (as one can
see in the explicit example in Ref.~\cite{Batell:2008zm}).

Third, one can also argue that in a theory with quarks, the energy of
a static string along $z$ direction at fixed $\vec x$ should be IR finite, since it is
related to the mass of the heavy-light meson. In the model
of~\cite{Andreev:2006ct} this energy diverges as $\int\!dz\,\tens(z)$.

Finally, the assumption of purely geometric background is too
simplistic. In a more realistic model, where the background arises
dynamically, one should expect other scalar fields, besides dilaton,
developing nontrivial profiles. These scalars could be expected to couple to
the string world sheet and thus affect the effective string tension $\tens(z)$.

Of course no one has ever constructed a consistent embedding of a soft
wall in string theory. As the modes that are captured by the low
energy effective action are excited modes of the flux tube and hence
the string such an embedding would necessarily be one where the
geometric features of the wall are string scale in size and so
presumably one would first have to understand string theory on
strongly curved spaces with RR flux before one could even hope to
construct a soft wall in string theory. But let us speculate a little
of what such construction should look like and this way give evidence
that the soft wall story with the inclusion of additional semi-classic
fundamental strings for external sources is more complicated

First of all, let us recall the standard wishful scenario of how a 5d
dual background to real QCD should come about, extrapolating from
theories whose holographic dual we do understand. It hard to say where
this story was first presented, but one early paper where important
aspects of this story are hinted at is \cite{Klebanov:1998yya}. In the
UV the geometry dual to QCD is almost AdS, capturing the conformal
invariance of the asymptotically free high energy limit. The string
dilaton, dual to the coupling constant of QCD, has some small flow
with $z$ corresponding to the logarithmic running of the coupling
which leaves its imprint on the geometry and leads to small deviation
from AdS. As we move deeper in the interior a second scalar field
comes into play, the closed string tachyon $T$. In the UV region the
tachyon was stabilized by the RR fluxes that supported the AdS
geometry, but the deformations in the geometry, triggered by the
running dilaton, allow it to recover its true nature. As the closed
string tachyon runs (as a function of $z$) to its true minimum we
enter a region where space and time cease to exist all fields are
frozen out, no dynamical degrees of freedom propagate in the tachyon
wall---the bulk manifestation of the confining gap. The important
ingredient here is that while the soft wall field $\Phi$ may still be
the dilaton, there are actually two scalar fields turned on. And both
of them couple to the fundamental string worldsheet (the closed string
tachyon plays the role of a cosmological constant on the worldsheet,
at least in the Polyakov formulation).

Let us present two supporting pieces of evidence. One is given by the
explicit realization of a soft wall in a simple toy model presented in
\cite{Batell:2008zm}. In our original work, we just took the soft wall
as a given background and did not ask for it to solve any particular
equation of motion itself. The authors of \cite{Batell:2008zm} managed
to find a very simple toy model Lagrangian that gave the soft wall as
a solution to Einstein's equations. As in our fairy tale, the
Lagrangian of \cite{Batell:2008zm} actually required two scalar
fields, which were roughly identified with string theory dilaton and
tachyon just as expected. The second piece of evidence comes from
thinking about mesons made out of heavy flavor. As argued recently in
\cite{Grigoryan:2010pj} it is very clear that a one-scale soft wall
model can not possibly account for heavy quark mesons which are
dominated by two independent scales, the quark mass and
$\Lambda_{QCD}$. In a stringy construction of the type we outlined
this is easily incorporated. The soft wall driving confinement is set
up by the closed string tachyon. This sets $\Lambda_{QCD}$. Heavy
flavor is included by a flavor D/anti-D branes similar to the
construction of \cite{Karch:2002sh}. This flavor brane system comes
with its own scalar field: the open string tachyon. Latter allows one
to end the D-branes at a position determined by the quark mass,
effectively introducing the 2nd scale. Both these constructions
strongly hint at the existence of a scalar sector with more than a
single scalar field.

Irrespective of the details of such speculations, there are two very crisp statements one can make:
\begin{itemize}
\item The static quark/anti-quark potential in a realistic soft-wall
  model is not a simple exercise of applying standard recipes, but
  always needs one more ingredient: an assumption about the coupling
  of the background scalar(s) to the string
  worldsheet. 
\item Without making any assumptions about the worldsheet action, one
  can apply a simple test of whether the model exhibits linear
  confinement: does it give rise to the correct Regge trajectories?
  The $a>0$ soft wall model, with higher-spin fields included, passes
  this test.
\end{itemize}

\section{Conclusion}

We elaborated on the statement in our earlier paper on the existence of an extra massless mode in a negative sign (meaning
$a<0$) soft wall.
We stand by our earlier statement that due to the appearance of this extra mode,
the negative sign wall needs to be discarded as unphysical. This conclusion can independently be reached by studying the higher
spin mesons. We explained that arguments based on statements about the potential between external test quarks should be seen as
a test of how the fundamental string has to be coupled to the soft wall background, not of the background itself and the meson
physics it describes. The most important issue that of course remains un-addressed by our analysis is that the soft wall model
is not a systematic approximation to any known holographic system. We hope that it will nevertheless continue to serve as an
interesting toy model.

\end{document}